\documentclass{PoS}

\usepackage{epsfig,times,color,cite,adjustbox,caption,wrapfig,lipsum,booktabs,comment}
\usepackage[numbers]{natbib}

\newcommand{\degs}{\mbox{\(^\circ \)}}

\title{Construction of a medium-sized Schwarzschild-Couder telescope
as a candidate for the Cherenkov Telescope Array: development of
the optical alignment system}

\ShortTitle{SC-MST as a candidate for the CTA: development of the optical alignment system}

\author{\speaker{D. Nieto}$^{1}$, S. Griffiths$^{2}$, B. Humensky$^{1}$, P. Kaaret$^{2}$, M. Limon$^{1}$, I. Mognet$^{3}$, A. Peck$^{3}$, A. Petrashyk$^{1}$,  D. Ribeiro$^{1}$, J. Rousselle$^{3}$, B. Stevenson$^{3}$, V. Vassiliev$^{3}$, P. Yu$^{3}$, for the CTA Consortium\footnote{Full consortium author list at http://cta-observatory.org}\\
{\footnotesize
\\$^{1}$ Columbia University, Department of Physics,
$^{2}$ Iowa University, Department of Physics and Astronomy,
$^{3}$ University of California Los Angeles, Division of Astronomy and Astrophysics.}

E-mail: \email{nieto@nevis.columbia.edu}}

\abstract{The Cherenkov Telescope Array (CTA) is an international project for a
next-generation ground-based gamma-ray observatory. CTA, conceived as
an array of tens of imaging atmospheric Cherenkov telescopes,
comprising small, medium and large-size telescopes, is aiming to
improve on the sensitivity of current-generation experiments by an
order of magnitude and provide energy coverage from 20 GeV to more
than 300 TeV. The Schwarzschild-Couder (SC) medium-size candidate
telescope model features a novel aplanatic two-mirror optical design
capable of a wide field-of-view with significantly improved imaging
resolution as compared to the traditional Davis-Cotton optics
design. Achieving this imaging resolution imposes strict alignment
requirements to be accomplished by a dedicated alignment system. In
this contribution we present the status of the development of the SC
optical alignment system, soon to be materialized in a full-scale
prototype SC medium-size telescope at the Fred Lawrence Whipple
Observatory in southern Arizona.}

\FullConference{The 34th International Cosmic Ray Conference,\\
		30 July- 6 August, 2015\\
		The Hague, The Netherlands}

\begin{document}
\setcitestyle{square}

\section{Introduction}

The Schwarzschild-Couder medium-sized telescope (SC-MST) is a
candidate telescope for the Cherenkov Telescope Array
(CTA\footnote{www.cta-observatory.org},~\cite{Acharya20133}), an
international endeavor towards the construction of the next-generation
of imaging atmospheric Cherenkov telescopes (IACTs). The SC-MST's
novel optics design offers a significantly better angular resolution
over a wider field of view (FoV) than conventional single-mirror
IACTs~\cite{2007APh....28...10V}. The SC-MST's realization implies new
technological challenges, like satisfying substantially more stringent
optics alignment tolerances.  This paper reports on the current
development status of the SC-MST alignment system, soon to be
materialized in a full telescope prototype unit (pSCT) which is now
under construction at the Fred Lawrence Whipple Observatory in
southern Arizona.

A more detailed description of the pSCT optical and mechanical systems
as well as its gamma-ray camera, along with a study on the SC-MST
potential contribution to the overall CTA performance can be found
elsewhere in these
proceedings~\cite{2015ICRC-OPT,2015ICRC-CAM,2015ICRC-MEC,2015ICRC-MC}.

This contribution is structured as follows: in
Section~\ref{sec:overview} we give a brief overview of the alignment
system; in Section~\ref{sec:gas} we describe the development status of
the global alignment system; the panel-to-panel alignment system is
dealt with in Section~\ref{sec:p2pas}; a short description of the
alignment system software control is given in
Section~\ref{sec:control}.

\section{Overview of the pSCT alignment system}
\label{sec:overview}

\begin{wraptable}{l}{0.5\textwidth}
   \centering
	\begin{adjustbox}{max width=0.5\textwidth}
   	\begin{tabular}{@{} lcc @{}}
      	\hline
      	&Primary mirror &Secondary mirror \\
      	\hline
      	Global alignment & Value & Value\\
      	\hline
      	Translation $\bot$ to optical axis& 10 mm & 10 mm \\
      	Translation $\parallel$ to optical axis& 17 mm & 5 mm \\
      	Tilt & 15 mrad & 0.15 mrad\\
      	\hline
      	Panel-to-panel alignment & Standard deviation & Standard deviation\\
     	\hline
     	Translation $\bot$ to optical axis&2.2 mm&1.1 mm  \\
      	Translation $\parallel$ to optical axis& 17mm & 4 mm  \\
	Rotation around tangent axis& 0.1 mrad& 0.2 mrad\\
	Rotation around radial axis&0.1 mrad & 0.3 mrad\\	
	Rotation around normal axis&16.2 mrad & 118 mrad\\
	\hline
       	\end{tabular}
   	\end{adjustbox}
   \caption{Independent transformations of the primary and secondary mirrors and their 
   mirror panels needed to increase the PSF size to 1 arcmin on axis. For
the panel-to-panel alignment, each one follows a Gaussian
distribution centered on the ideal position.
   }\label{tab:alignment_req}

\end{wraptable}

The two-mirror SC optical system is designed to fully correct
spherical and comatic aberrations while providing a large field of
view and a fine plate-scale, allowing for finely pixelized focal plane
instrumentation. To optimize production costs, the 9.7 m diameter
primary mirror (M1) and the 5.4 m diameter secondary mirror (M2) are
segmented into 48 and 24 mirror panels respectively. Extensive
ray-tracing simulations of the SC-MST optical system indicated the
need for sub-millimeter and sub-milliradian precision for both global
and panel-to-panel alignments in order to achieve a point spread
function (PSF) compatible with the pixel size of a high-resolution
gamma-ray camera (see Table~\ref{tab:alignment_req}). In addition, the
accuracy of source localization requires 5 arcsec mirror tilt
control. These tolerances are roughly equivalent to a sub-mm radio
telescope operating in the range 100 - 20 $\mu$m and three to four
orders of magnitude above the usual diffraction limit of optical
telescopes. Nevertheless, these tolerances are far more demanding than
those of current IACT optical systems and require automated mechanical
alignment of the segmented mirror surfaces.

The global alignment system is designed to continuously measure
relative positions of the main optical elements of the telescope, M1,
M2 and the camera focal plane, as well as to detect large-scale
spatial perturbations of the M1 and M2 figures. The panel-to-panel
alignment system is designed to effectively detect and correct for
misalignment between neighboring panels as well as to continuously
monitor the alignment of the optical surfaces' figures. The
integration of both systems will ensure the aforementioned
sub-millimeter and sub-milliradian precision in the alignment and
positioning of the pSCT's main optical elements.

\section{Global alignment system}
\label{sec:gas}

\begin{figure*}[!t]
  \centering	
  \includegraphics[width=0.4\linewidth,clip=true,trim= 0 40 0 40]{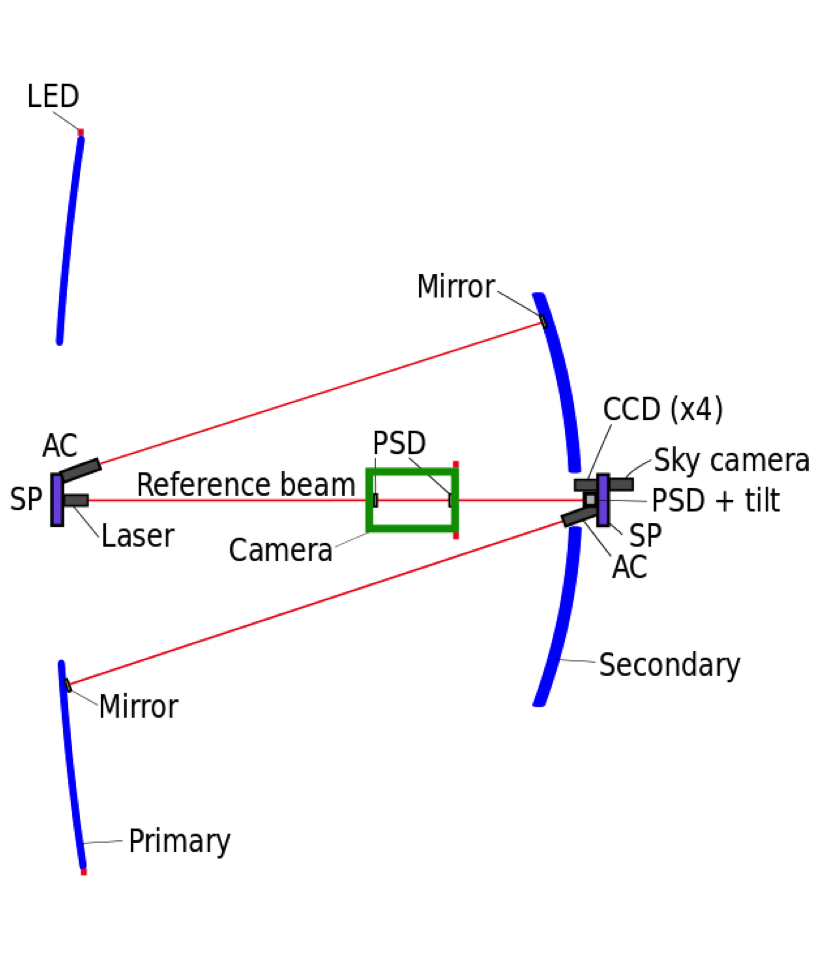}
  \includegraphics[width=0.5\linewidth]{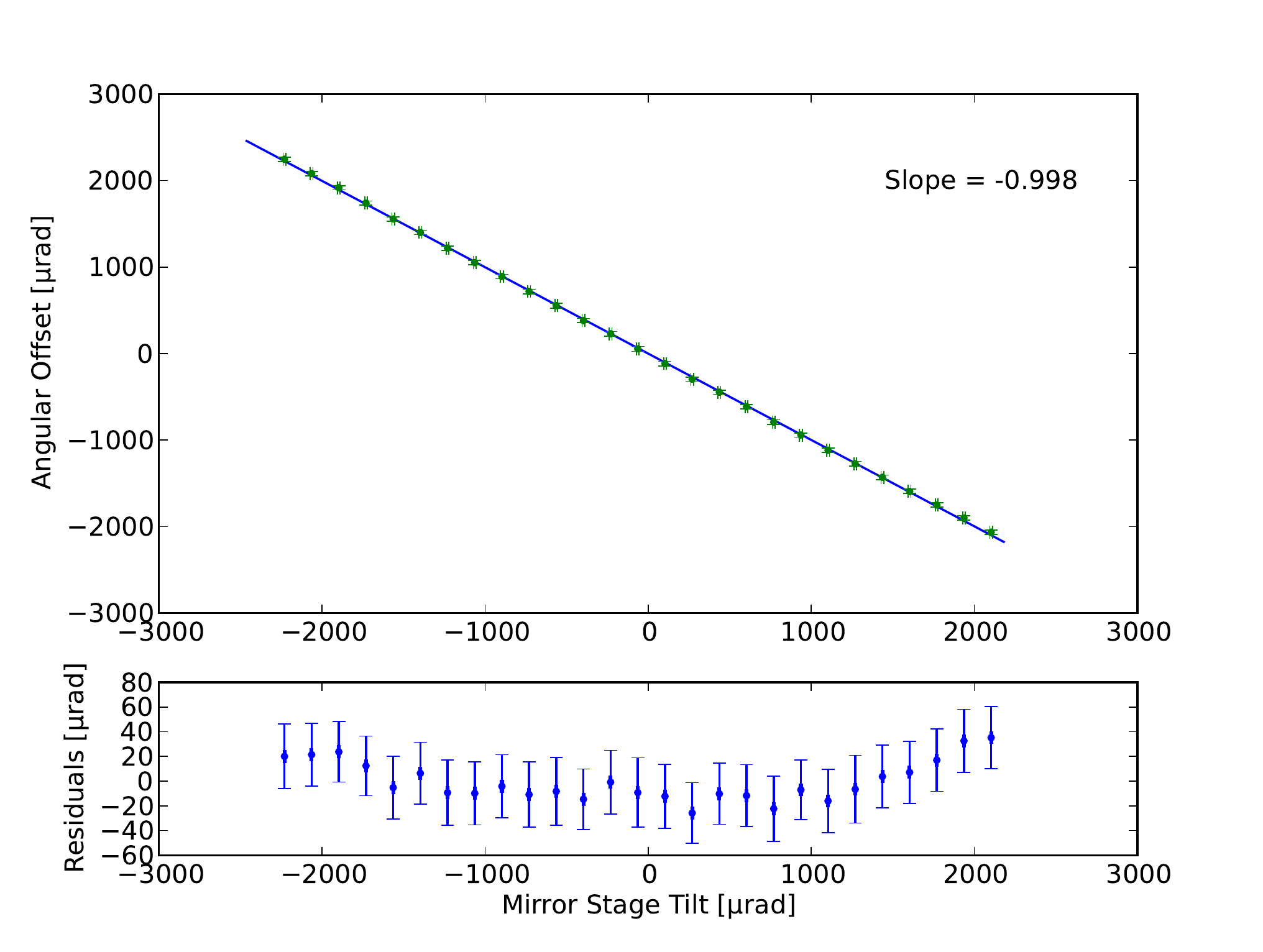}
  \caption{{\em Left}: Schematic description of the global alignment subsystem showing all relevant devices. AC stands for autocollimator, SP stands for Stewart platform, and PSD stands for position-sensitive device. {\em Right}: Reconstructed tilt from a prototype autocollimator placed 8~m from a 5~cm circular mirror mounted on a tilt stage.} 
  \label{fig:ga_ccdled}
\end{figure*}

The global alignment system for the pSCT consists of CCD cameras
imaging LEDs to measure the relative translations between M1, M2, and
the gamma-ray camera, and autocollimators to measure the tilts of M1
and M2 (see Fig.~\ref{fig:ga_ccdled}, left panel, for a schematic
view).  Prototypes of all of the components have been built and
tested.

The performance of the translation measurement system was tested as
follows. A set of six LEDs were arranged on a panel in a configuration
similar to what will be used on the pSCT.  The LEDs were imaged with a
CCD camera, roughly 8 meters away, which was moved along the line
between the LED panel and the camera.  An image was acquired at each
position and the position of the LED panel was reconstructed from the
known relative positions of the LEDs and their positions as measured
on the image.  The reconstruction uses a forward folding technique to
iteratively derive the translation and rotation that provide the best
fit to the LED positions measured on the image.  We note that the
focal length of the camera's lens is a critical parameter in the
reconstruction.  The Fujinon lenses used have a manufacturing
tolerance of 5\%, so we measured the lens focal length.  A temperature
sensor will be mounted on each camera to allow correction of any
temperature dependence of the focal length.

The scatter in the known versus reconstructed positions gives an
indication of the accuracy of the system.  After a linear fit to the
data, residuals ranging from 0.6~mm to 1.3~mm (rms) were found.  This
is well within the accuracy required for the measurement of the mirror
panel positions along the optical axis of the telescope.  The
residuals for the orthogonal translations were 0.11~mm, again, well
within the required accuracy.  This system also allows measurement of
the tilt of the mirror panels to an accuracy of about 1.1~mrad.  This
in insufficient to reach the CTA requirements on accuracy of source
locations and we employ autocollimators to measure the mirror panel
tilts.

Due to cost constraints, we chose to build autocollimators instead of
buying commercial units.  Instead of the standard autocollimator
design that uses a laser sensed by a position-sensitive photodiode, we
developed a design in which light from an LED is projected through an
achromatic doublet onto a mirror and then the returning light is
imaged with a CCD camera.  This system can tolerate larger mirror
misalignment than commercially available autocollimators.

We performed tests with the autocollimator placed 8~m from a 5~cm
circular mirror mounted on a tilt stage.  We tilted the mirror over a
range of $\pm 2000~\mu$rad and reconstructed the tilt from the
reflected image of the LEDs obtained with the camera. The right panel
on Fig.~\ref{fig:ga_ccdled} summarizes the results from these tests,
where we found an rms deviation from a linear fit of 16~$\mu$rad. This
is within the design goals.

\section{Panel-to-panel alignment system}
\label{sec:p2pas}

The positioning of each individual mirror panel is achieved by six
linear actuators arranged in a Stewart platform (or hexapod) designed
to situate the mirror panel with an accuracy better than the required
100~$\mu$m. The relative positions of adjacent mirror panels are
determined by a collection of mirror panel edge sensors (MPESs). The
control of the SP and the MPESs is performed by a dedicated mirror
panel controller board (MPCB) mounted onto an aluminum triangle that
interfaces the entire setup with the optical support structure
(OSS). We define a panel module (PM) as the full setup composed of a
mirror panel, and its accompanying SP, MPES, MPCB and mounting
triangle. Power and Ethernet are served to each of the 48 PMs in M1
and 24 PMs in M2 by two power and Ethernet distribution boxes (PEDBs),
one per mirror. In turn, all the 72 PMs are controlled by a central
computer in charge of the pSCT alignment.

\subsection{Mirror Panel Edge Sensors}

\begin{wrapfigure}{L}{0.5\textwidth}
  \centering
  \includegraphics[height=4cm]{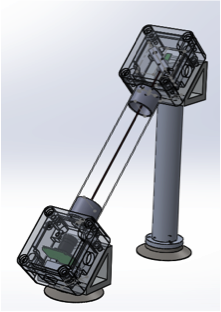}
  \includegraphics[height=4cm,clip=true,trim= 100 50 150 20 ]{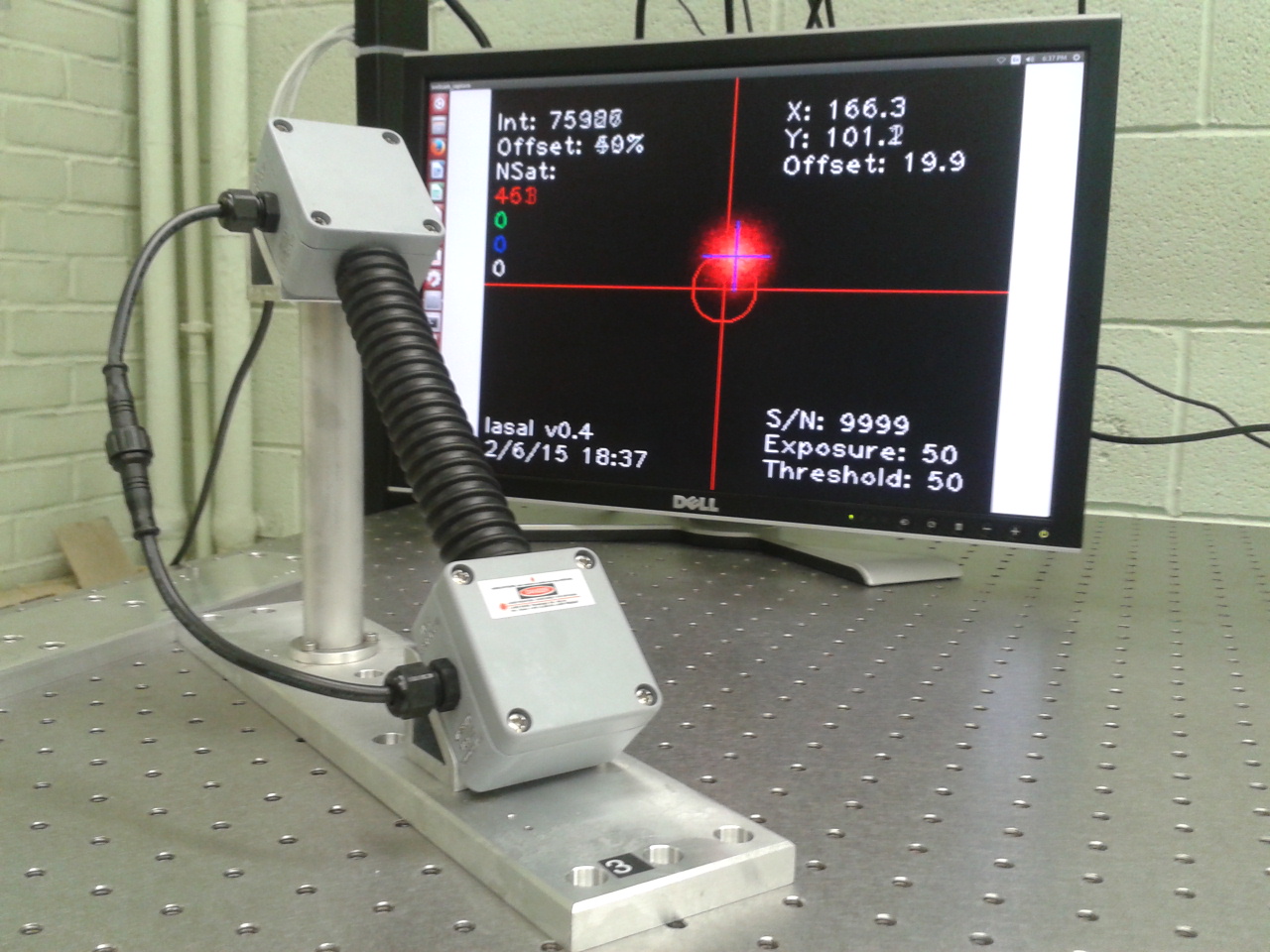}
  \caption{{\em Left}: CAD model of a MPES unit showing the
    localization of the webcam and the laser. {\em Right}: Actual MPES
    unit mounted on a test platform. A real-time reading and analysis
    of the MPES image can be found in the background monitor.}
  \label{fig:MPES}
\end{wrapfigure}

The MPESs are required to provide a positional resolution better than
10 $\mu$m over an operational area of $\sim$1 cm$^{2}$. The MPES
measurements will be used to initially align the pSCT optical
surfaces, providing input to compute the necessary adjustments of the
SPs on individual mirror panels to compensate potential deformations
of the OSS. These potential deformations may be caused by seasonal
thermal expansion and/or changing loads in the pSCT structure as it
moves. Although adjustments of the PMs are expected to be infrequent,
MPESs will be performing measurements continuously to accumulate data
for pointing calibration and PSF estimation.

Each MPES unit consists of two mechanically independent subunits: the
sensor subunit and the light source subunit, as shown in
Fig.~\ref{fig:MPES}.  Each subunit consists of a weatherproof
enclosure, containing the optical and electronic components. The light
source subunits features a low-power diode laser whose beam is further
collimated by a 300~$\mu$m diameter pin hole located at the exit of
the enclosure. The sensor subunit features a 0.3 Mpx webcam focused
onto an opal glass that works as a diffusive target for the
perpendicularly incident laser beam. Both subunits are connected and
powered by the same USB line that is used to communicate with the
webcam. A removable opaque UV-resistant thermoplastic elastomer
conduit connects both subunits forming a single MPES unit, in such a
way that the laser beam is light-tight and weatherproof.  The whole
MPES design has been proven to be weatherproof and has successfully
passed total water submersion tests. It has been demonstrated through
extensive laboratory tests that a < 5 $\mu$m positional resolution
with a plate scale of 44$\mu$m/px can be achieved with the current
MPES design.

Each MPES presents a single optical axis defined by its laser
beam. There is a single MPES configuration where the optical axis
orientation forms an angle of $\sim$45$\degs$ with the local mirror
panel surface. Rotations of this basic configuration allows for triads
of MPES with orthogonal optical axes. Adjacent mirror panels within
the same ring segment will be interfaced with a triad of MPESs,
consequently optimizing the measurement of the displacements in the
allowed 6 dimensional space. Adjacent mirror panels belonging to
different ring segments share a pair of MPES, thus allowing for
inter-segment displacement measurements and for redundancy. The total
number of MPESs per pSCT provides a 20\% redundancy that should not
cause malfunctioning of the alignment system in the event of
malfunctioning of a small group of units. The installation of a
limited number of MPESs interfacing the optical surfaces with the OSS
is under consideration.

The production, alignment and calibration of circa 400 MPES units is
currently ongoing. All units will be manufactured on time for the
integration into the pSCT PMs.

\subsection{Stewart Platforms}

The actuators configuring the SP are capable of stepping in increments
of 3 $\mu$m within a range of 63 mm, while their position is measured
by a magnetic encoder capable of detecting a single step. Both
stepping motor and encoder are enclosed in a watertight aluminum
cylinder.

The actuators are connected on both sides of the SP to six universal
joints which are designed to minimize hysteresis using precisely
machined aluminum and a crossed-axis configuration. The addition of
the joints and actuators provide six degrees of freedom to move the
mirror panels in any position within the range of the stepping motor
(63 mm). The resulting SP occupies a space of approximately
$600\times600\times600$ mm$^3$. A stepping motor alone is capable to
move a maximum weight of 65 kg at 500 steps/s. After assembly, the SP
is able to move at least 50 kg in addition to its own weight ($\sim$
15 kg), more than enough to hold and align a single mirror panel ($<$
25 kg).

\begin{figure*}[!t]
  \centering
  \includegraphics[height=3.7cm]{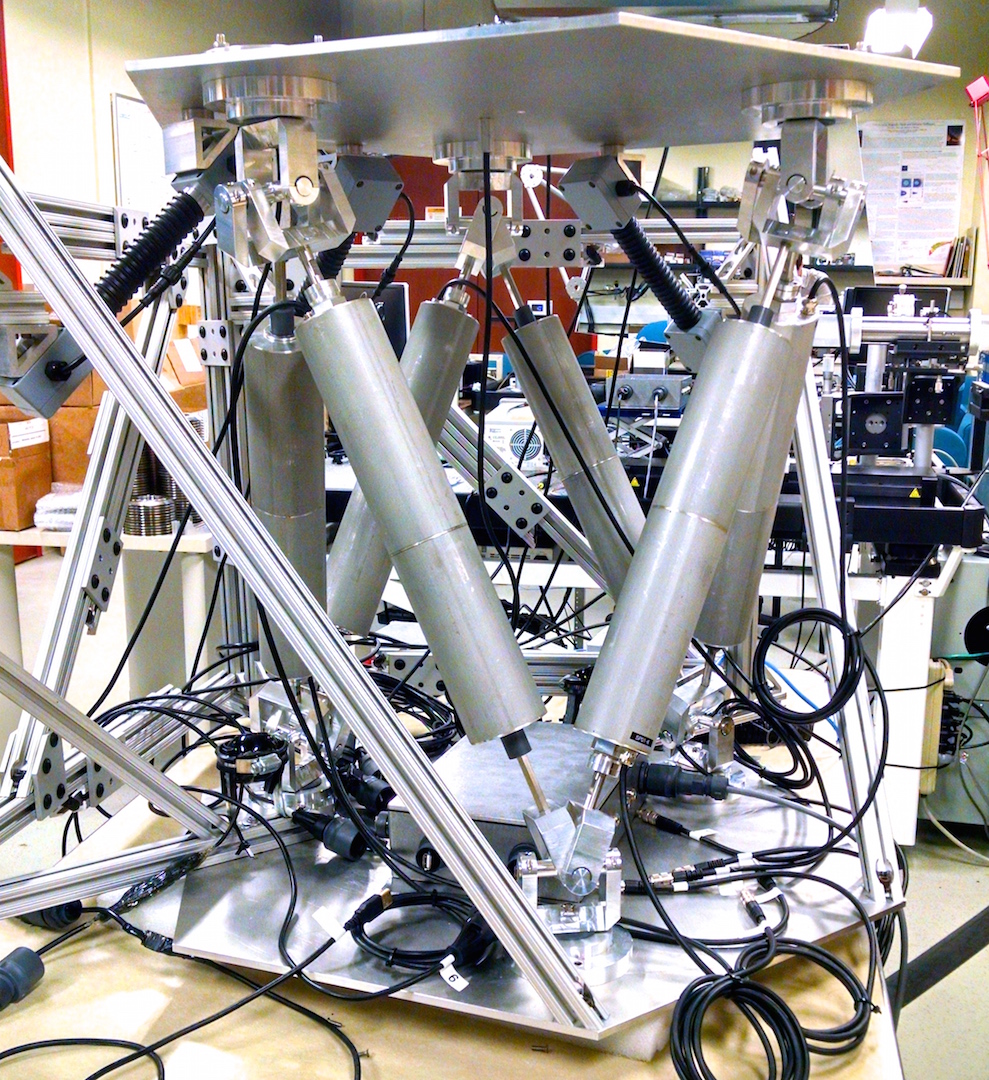}
  \includegraphics[height=4.2cm,clip=true,trim=0 30 0 0]{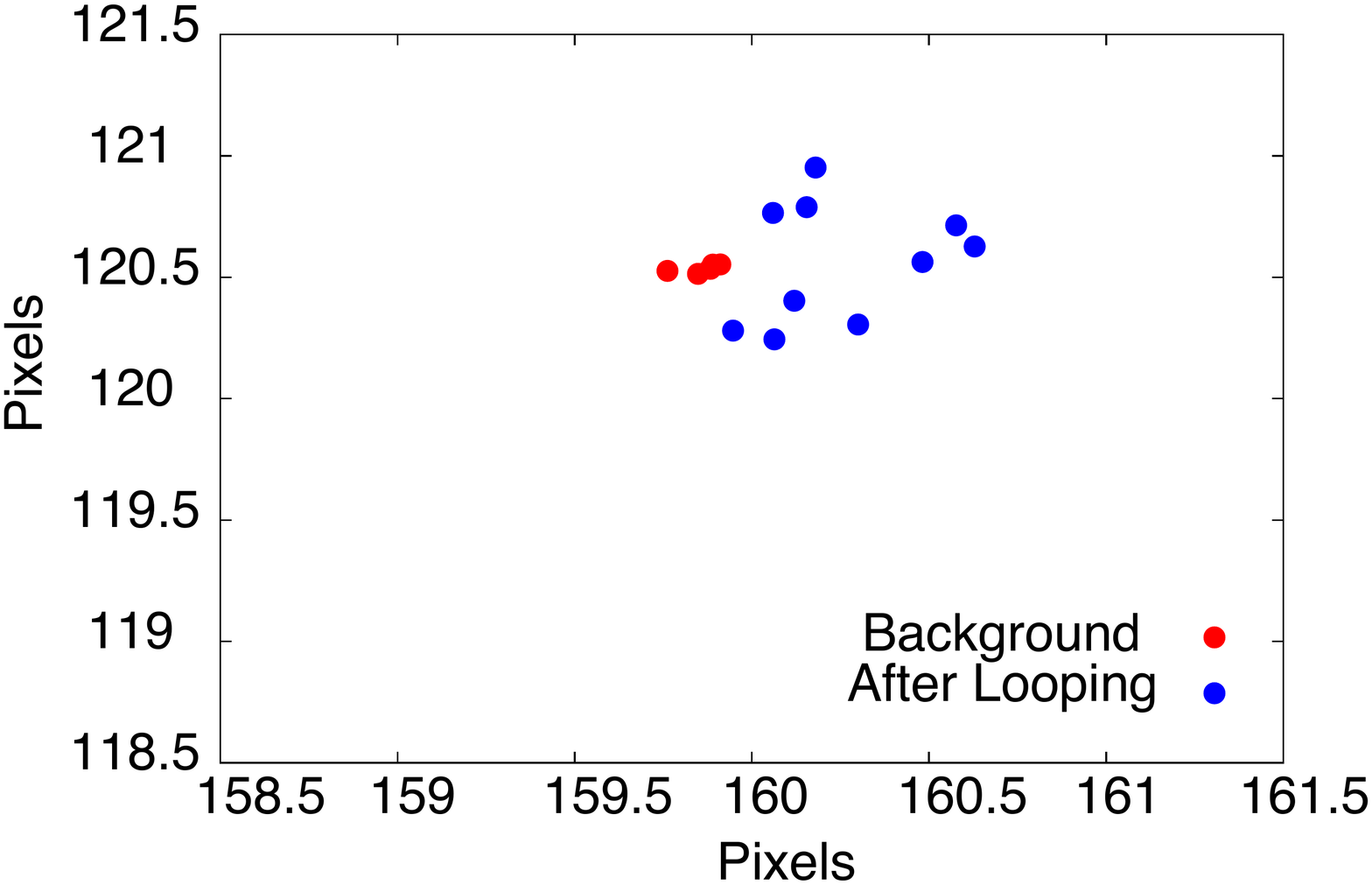}
  \includegraphics[height=4.2cm,clip=true,trim=0 30 0 0]{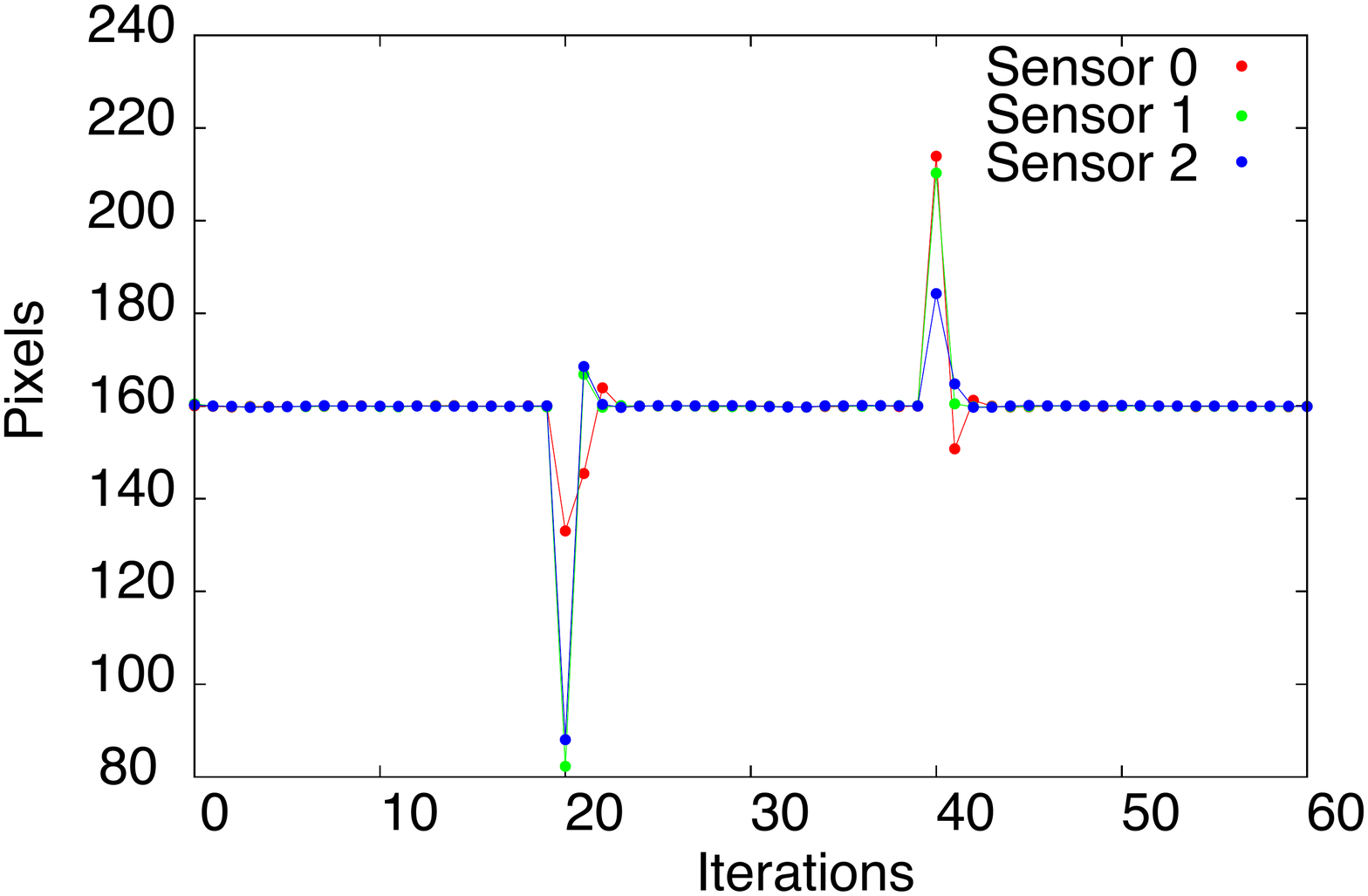}
  \caption{{\em Left}: Laboratory setup used to measure the hysteresis
    of the Stewart platform, including actuators and three MPES. {\em
      Center}: Example of hysteresis measurement. The red dots
    represent the MPES positions in pixels (44~$\mu$m/px) while the
    platform is not moving. The blue dots show the MPES positions
    after moving the platform back and forth through different
    sequences of actuator motions. {\em Right}: MPES position in pixel
    coordinates. A perturbation is introduced to the platform position
    which is corrected by a feedback loop between actuators and MPES
    in 3 iterations.}
  \label{SP}
\end{figure*}

The 3~$\mu$m steps of the actuators should provide a tip-tilt angular
resolution of 1 arcsec for the mirror panels, but the actual alignment
precision of the SP is mainly limited by the remaining hysteresis of
the joints. This effect has been measured in the laboratory using
prototype versions of the actuators and MPES, which were used to
monitor the position of the SP (Fig.~\ref{SP}, central panel). The
stability of this setup, which was measured using the MPES while the
SP remains immobile, is better than 10~$\mu$m (see Fig.~\ref{SP},
central panel, red dots). The effect of hysteresis was measured by
moving the SP to a new position using a specific sequence of the
actuators motion and moving back to the initial position through a
different sequence. As shown in Fig.~\ref{SP} (central panel, blue
dots) the SP does not reach the exact same initial position, and the
spread of these positions quantifies the amount of hysteresis
remaining in the system. The measured standard deviations of these
distributions are smaller than 26 $\mu$m, depending on the load on the
platform and its intermediate positions.

It is possible to mitigate some of the effects of hysteresis on the SP
by re-adjusting several times its position around a desired value
using a feedback loop with the MPES. Fig.~\ref{SP} (right) shows some
of the results obtained in laboratory where the actuators are
constantly adjusted to reach a given position after inserting
artificial perturbations. After perturbation, the SP was able to
converge to the desired position within three iterations and to
maintain it within 15 ~$\mu$m.  To achieve this result the response
matrix of the actuators has been measured with the MPES using a linear
approximation. This matrix was then inverted and applied several times
to the actuators to reach a specific position on the MPES.

These results were obtained using a late prototype of the SP shown in
Fig.~\ref{SP} (left), which was used to produce the 72 platforms
needed for the pSCT. All the parts of these platforms have been
procured and are currently being assembled to be installed on the OSS
around Fall 2015.

\subsection{Mirror Panel Controller Board}

Each PM will hold a controller box behind it. This MPCB is responsible
for driving the actuators and the collection of the signal coming form
the MPES, magnetic encoders and temperature sensors. The signal
processing, communication and control will be handled by a Gumstix
Overo$^{\tiny \textregistered}$~EarthSTORM on-board micro computer
with Ethernet connectivity.

The production of 80 controller boxes is complete, and they are
awaiting final testing and calibration, followed by installation of
the production software.

\subsection{Power and Ethernet Distribution Box}

The PEDB system has two main functions: distribute Ethernet and 24 VDC
power to all the telescope's PMs. Additionally, the boxes supply power
(12 VDC and 24 VDC) and Ethernet to several other systems including
the global alignment system, and the gamma-ray camera LED flashing
calibration system. Both M1 and M2 will be equipped with one PEDB each
that will be installed on the OSS. Both PEDB units are housed in NEMA
3R outdoor-rated enclosures with ported forced-air cooling and
temperature control. Ethernet connectivity is provided by two
commercial-grade switches. Housekeeping and power supply control is
provided by an Arduino Mega 2560 with Arduino$^{\tiny \copyright}$
Ethernet shield. Additional analog input lines are provided by custom
boards containing 12 bit ADCs which communicate via serial peripheral
interface to the Arduino$^{\tiny \copyright}$. All power supply output
voltages are monitored, as well as the currents to all Stewart
platforms.

The two PEDBs for installation on the pSCT are currently undergoing
final electrical testing and final installation of remaining
connectors. Both boxes have been run for extensive periods in the
laboratory and have not exhibited any abnormal behavior.

\subsection{Integration and environmental tests}

\begin{figure*}[!t]
  \centering
  \includegraphics[width=0.24\textwidth,clip=true,trim=0 0 50 50]{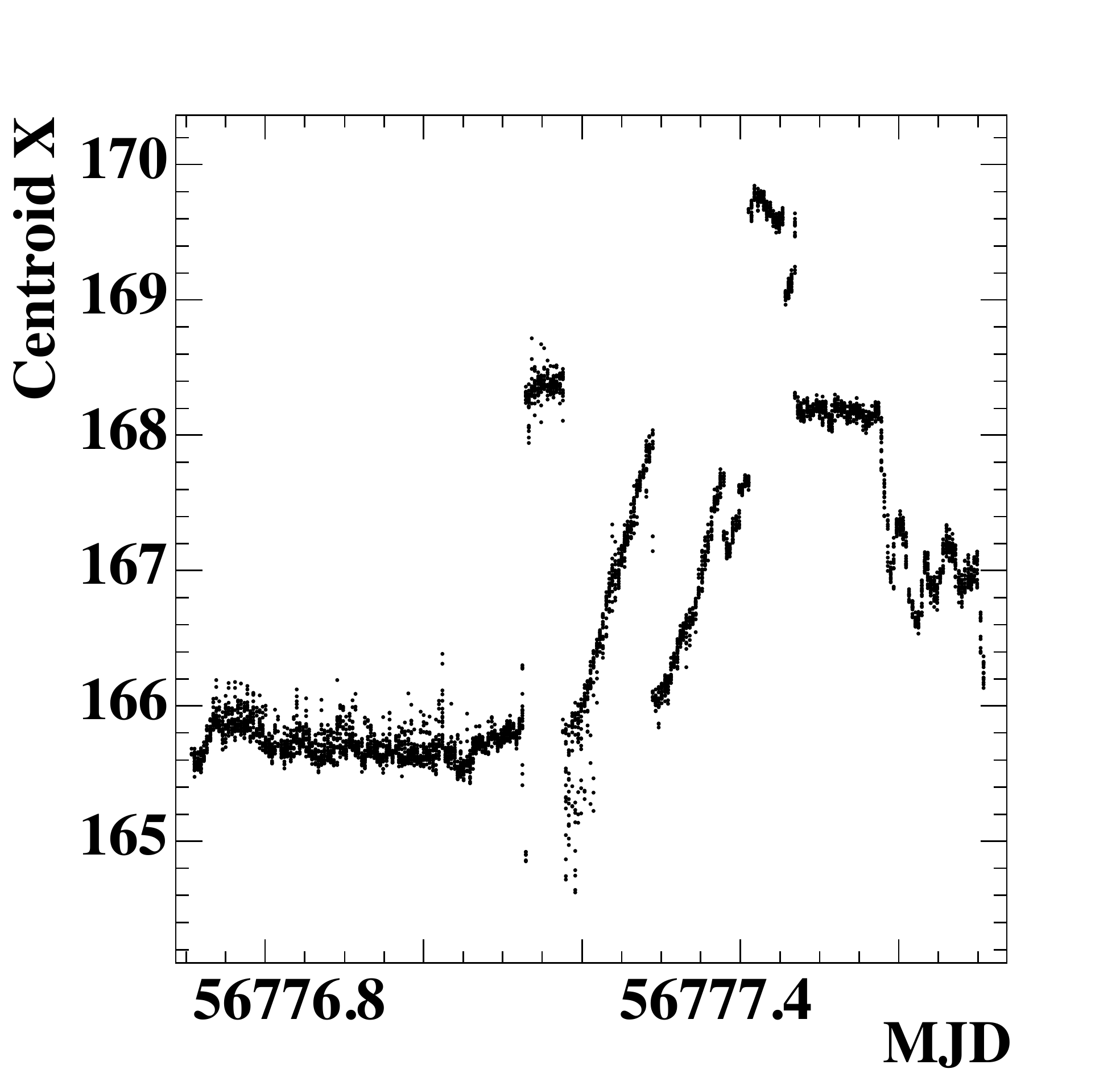}	
  \includegraphics[width=0.24\textwidth,clip=true,trim=0 0 50 50]{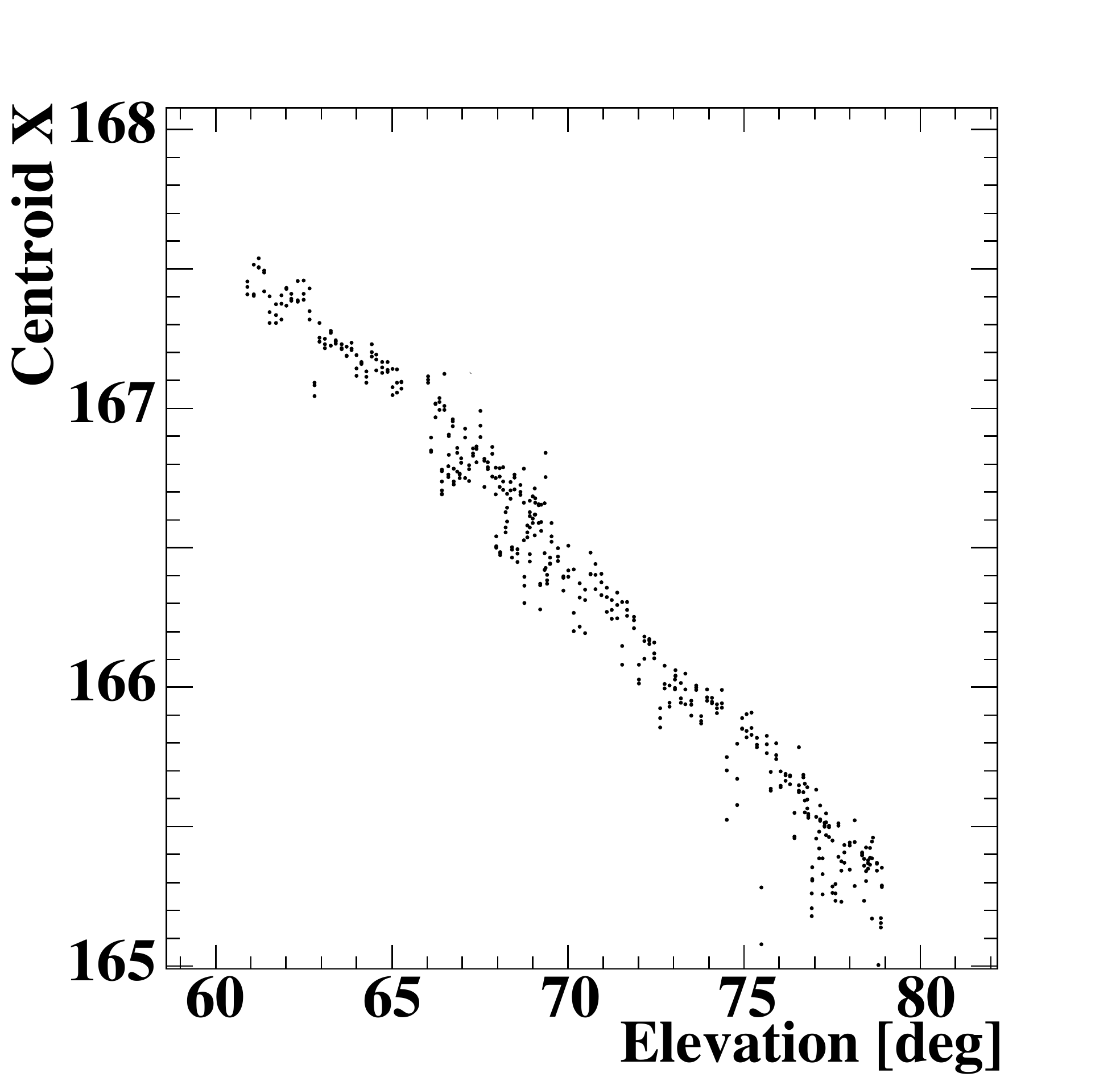}
  \includegraphics[width=0.24\textwidth,clip=true,trim=0 0 50 50]{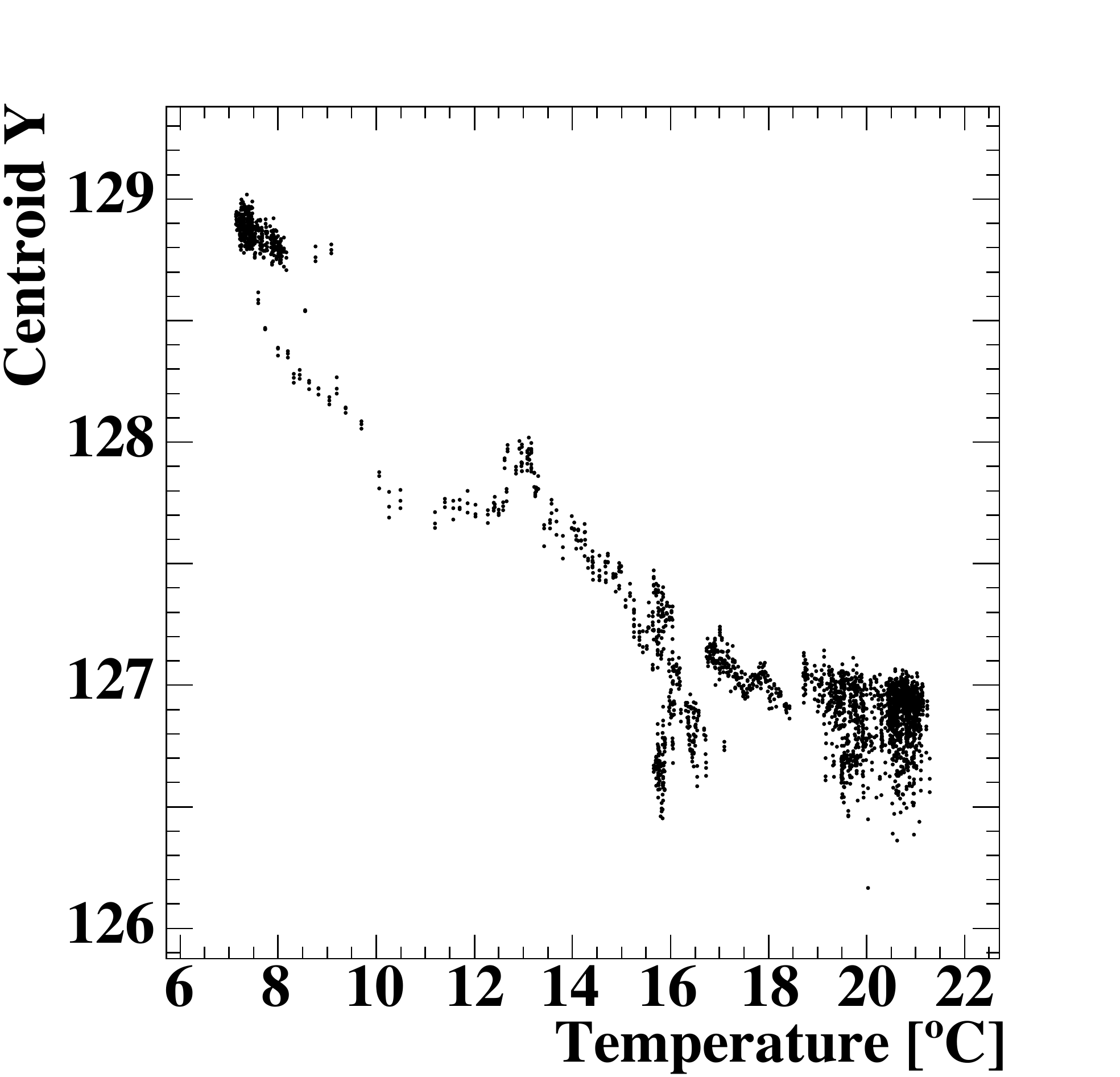}
  \includegraphics[width=0.24\linewidth,clip=true,trim= 150 0 950 0]{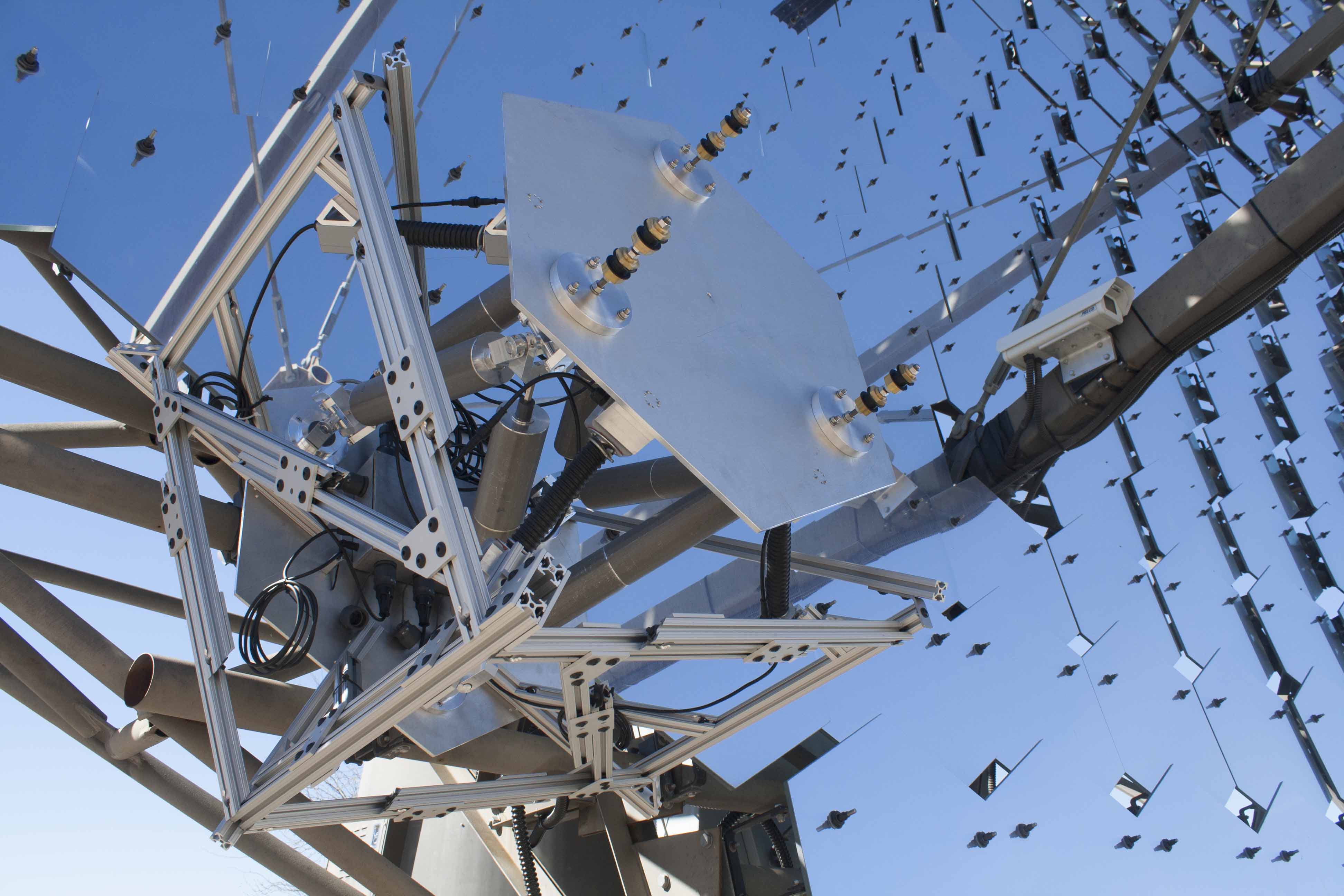}
  \caption{Field test results: readings from a single MPES over a 24h
    period. {\em Right} evolution of the centroid X position; {\em
      center right}: centroid X position drift as a function of
    VERITAS T4 elevation during observing time; {\em center left}:
    centroid Y position drift as a function of environment temperature
    at VERITAS T4 stow position; {\em left}: pPM as installed on
    VERITAS T4.}
  \label{fig:mpes_fieldtest_results}
\end{figure*}

A panel module prototype (pPM) consisting of a full SP, a MPCB, an
orthogonal triad of MPESs and a dummy aluminum-made mirror panel,
along with a PEDB prototype, were assembled at the Fred Lawrence
Whipple Observatory and installed on one of the VERITAS telescopes as
a first hardware integration test. The setup, which ran continuously
for 4.5 months starting in early 2014, demonstrated that the MPESs'
positioning measurements and the SP motion control were both within
specifications, as well as positively verified the weatherproofness of
all the participating hardware.

The MPESs measured relative true motion of the dummy aluminum plate
correlated with changes in the gravitational loading of the assembly
as the telescope slewed, as well as the relative true motion
correlated with ambient temperature changes that is attributed to the
thermal expansion of the pPM. Some examples of data recorded by an
individual MPES can be found in Fig.~\ref{fig:mpes_fieldtest_results}
along with a picture of the pPM.

\section{Alignment control software}
\label{sec:control}

The software controls of the panel-to-panel alignment system are
implemented through the OPC-UA communication protocol. The
implementation follows the server-client paradigm, with MPCBs on each
panel acting as servers of MPES data; PEDBs on each mirror being
servers of the system environment data; a central computer fulfilling
the roles of client to the aforementioned servers and server of the
processed data; and a database collector, a client to the central
computer that handles archival data requests between the central
computer and a database. The global alignment system control software
will be similarly implemented, such that the entire pSCT alignment
control software easily integrates with the higher-level Array Control
system of the CTA.

\section{Summary and outlook}
\label{sec:summary}

We have presented an overview of the design and development status of
the alignment system for a Schwarzschild-Couder medium-sized telescope
candidate for the Cherenkov Telescope Array. The performance of all
hardware components for the global and panel-to-panel alignment
systems has been verified to work within specifications and the mass
production of the same has been initiated and is compliant with the
pSCT construction schedule.

For the final SC-MST design, the alignment system will be optimized
for mass-scale production based on the knowledge acquired from the
pSCT experience.

\section{Acknowledgments}

We gratefully acknowledge support from the agencies and organizations
listed under Funding Agencies at this website:
http://www.cta-observatory.org/. The development of the prototype SCT
has been made possible by funding provided through the NSF-MRI
program. We acknowledge the excellent work of the technical support
staff at the Fred Lawrence Whipple Observatory.

{
\small
\setlength{\bibsep}{0pt}
\bibliography{alignment}{}

\providecommand{\href}[2]{#2}\begingroup\raggedright\begin{thebibliography}{1}

\bibitem{Acharya20133}
B.~Acharya et~al., {\it Introducing the {CTA} concept},  {\em Astroparticle
  Physics} {\bf 43} (2013), no.~0 3 -- 18.

\bibitem{2007APh....28...10V}
V.~V. {Vassiliev}, S.~{Fegan}, and P.~{Brousseau}, {\it {Wide field aplanatic
  two-mirror telescopes for ground-based {$\gamma$}-ray astronomy}},  {\em
  Astroparticle Physics} {\bf 28} (Sept., 2007) 10--27,
  [\href{http://arxiv.org/abs/astro-ph/0612718}{{\tt astro-ph/0612718}}].

\bibitem{2015ICRC-OPT}
J.~{Rousselle} et~al., {\it {Construction of a Schwarzschild-Couder telescope
  as a candidate for the Cherenkov Telescope Array: Implementation of the
  optical system}},  {\em Proceedings of the 33rd International Cosmic Ray
  Conference} (2013) [\href{http://arxiv.org/abs/1307.4072}{{\tt
  arXiv:1307.4072}}].

\bibitem{2015ICRC-CAM}
N.~{Otte} et~al., {\it {Development of a SiPM camera for a Schwarzschild-Couder
  Cherenkov telescope}},  {\em 34th International Cosmic Ray Conference}
  (2015).

\bibitem{2015ICRC-MEC}
K.~{Byrum} et~al., {\it {A medium sized Schwarzschild-Couder Cherenkov
  telescope design proposed for the Cherenkov Telescope Array}},  {\em
  Proceedings of the 34th International Cosmic Ray Conference} (2015).

\bibitem{2015ICRC-MC}
T.~{Hassan} et~al., {\it {Layout design studies for medium-size telescopes
  within the Cherenkov Telescope Array}},  {\em Proceedings of the 34th
  International Cosmic Ray Conference} (2015)
  [\href{http://arxiv.org/abs/1508.06076}{{\tt arXiv:1508.06076}}].

\end{thebibliography}\endgroup
\bibliographystyle{JHEPs}
}
\end{document}